\documentclass[12pt,reprint,aps,groupedaddress,showpacs,nofootinbib,prl,showkeys]{revtex4-1}
\usepackage{amsmath,amssymb,graphicx,amsfonts}
\usepackage{mathrsfs}

\usepackage{hyperref}
\usepackage{setspace}

\begin{document}

\title{A family of solutions to the inverse problem in gravitation: building a theory around a metric}


\author{Arthur G. Suvorov}
\email{arthur.suvorov@tat.uni-tuebingen.de}
\affiliation{Theoretical Astrophysics, IAAT, University of T{\"u}bingen, T{\"u}bingen 72076, Germany}

\date{\today}

\begin{abstract}
\noindent{A method is presented to construct a particular, non-minimally coupled scalar-tensor theory such that a given metric is an exact vacuum solution in that theory. In contrast to the standard approach in studies of gravitational dynamics, where one begins with an action and then solves the equations of motion, this approach allows for an explicit theory to be built around some pre-specified geometry. Starting from a parameterized black hole spacetime with generic, non-Kerr hairs, it is shown how an overarching family of theories can be designed to fit the metric exactly.}
\end{abstract}


\maketitle

\section{Introduction}
In the study of Lagrangian field theory and the calculus of variations, one typically begins with an action functional and then investigates the dynamics of the associated theory. In many cases however, the inverse problem is also of fundamental interest \cite{doug41,inv1}: starting from a particular field configuration, can one find an invariant Lagrangian density whose equations of motion admit that field as an exact solution? Owing to the complexity of the differential equations involved, which are typically non-linear in realistic problems, finding such a Lagrangian, let alone all Lagrangians, can be a challenging task. This is especially true in studies of gravitation (e.g., \cite{grav1,grav2}), where the action is built from geometric invariants which depend on the tensorial metric field in complicated ways. Even conceptually simple cases like the $f(R)$ theories \cite{felice}, which involve only some function, $f$, of the Ricci scalar curvature, $R$, admit rich configuration spaces \cite{clifbar,fr2,suv20}. Nevertheless, substantial progress has been made on the gravitational inverse problem from cosmological observations in recent years \cite{od1,od2,od3}.

In the context of tests of general relativity (GR) from observations of compact objects, two main techniques are employed. One approach (sometimes called `top-down') involves picking a particular theory of gravity and comparing the solutions obtained within that theory with a suitable GR counterpart (e.g., \cite{top1,top2}). In this way, the predictions of a given theory are challenged directly using experimental data. Top-down methods are however limited because exact solutions describing realistic compact objects within a given theory are often not known, and it can be impractical to test multiple theories simultaneously using a given framework. The other approach (`bottom-up') involves a phenomenological parameterization of the spacetime that incorporates generic deformations of the GR counterpart \cite{nkerr1,nkerr2,nkerr3,contfrac}. However, even if the deviation parameters of the parameterized metric can be constrained, bottom-up approaches do not necessarily guide one towards the `true' theory of gravity. Moreover, backreaction effects cannot be self-consistently accounted for when a metric is considered independently of a parent theory \cite{bar08,suv19}. A unification of these two approaches, which would remedy the above shortcomings, boils down to requiring a solution to the inverse problem: given a metric (reconstructed from astrophysical data), find a (theoretically robust) theory of gravity that supports the solution exactly.

In this Letter we show how one can construct such a theory. In particular, a class of mixed scalar-$f(R)$ theories are presented which involve a function $f$ whose argument includes the scalar curvature and both potential and kinetic terms of a scalar field in a precise way. We show that there are large families of functions $f$ such that, for a particular scalar-field configuration, a given metric is an exact solution to the equations of motion. The theory reduces to a number of well-known cases in some limits, though in general has such a large configuration space so as to encompass practically any metric for some $f$. While the presented theory provides only one particular (not necessarily physically-motivated) example of a covariant action that can be tailored to a given metric, having an explicit construction on hand helps toward finding a general solution to the inverse problem. {The approach has the benefit that gravitational perturbations of a given spacetime can be accounted for self-consistently when using bottom-up methods. This could lead to improved accuracy in studies involving extreme-mass-ratio inspirals (EMRIs) \cite{nkerr1,emri2}, magnetohydrodynamical models of accretion disks \cite{joh16,bam17}, or quasi-normal ringing \cite{fran19,vol20} that aim to test GR but treat backreaction approximately.}

Except where needed for clarification, natural units with $c=G=1$ are adopted throughout.

\section{A mixed scalar-f(R) gravity}
Consider the theory governed by the action
\begin{equation} \label{eq:genscalarten}
\mathcal{A} = \kappa \int d^{4} x \sqrt{-g} f \big( F (\phi) R + V (\phi) - \omega ( \phi ) \nabla_{\alpha} \phi \nabla^{\alpha} \phi \big),
\end{equation}
where $\kappa = \left( 16 \pi G \right)^{-1}$, $G$ is Newton's (bare) constant, $R \equiv R_{\mu \nu} g^{\mu \nu}$ is the scalar curvature for metric tensor $g$, and $F$, $V$, and $\omega$ are potential functions of the scalar field $\phi$. When the function $f$ is linear in its argument $X$, where $X \equiv F(\phi) R + V(\phi) - \omega(\phi) \nabla_{\alpha} \phi \nabla^{\alpha} \phi$, the theory described by the action \eqref{eq:genscalarten}, which is a member of the general class considered by Hwang and Noh \cite{hwang1,hwang2}, reduces to standard scalar-tensor theory in the Jordan frame \cite{fujii,damour92}. The $f(R)$ theory of gravity is also recovered for constant scalar field {and vanishing potential $V$} \cite{felice}. {Here we consider only vacuum solutions of this theory, although matter fields could be included in the usual way.}

The equations of motion for the theory \eqref{eq:genscalarten} are found via the Euler-Lagrange equations, and are qualitatively similar to those of $f(R)$ gravity. Variation of \eqref{eq:genscalarten} with respect to the metric yields
\begin{equation} \label{eq:field1}
\begin{aligned}
0 =& F(\phi) f'(X) R_{\mu \nu} - \frac {f(X)} {2} g_{\mu \nu} + g_{\mu \nu} \square \left[ F(\phi) f'(X) \right] \\
& - \nabla_{\mu} \nabla_{\nu} \left[ F(\phi) f'(X) \right] - \omega(\phi) f'(X) \nabla_{\mu} \phi \nabla_{\nu} \phi,
\end{aligned}
\end{equation}
while variation with respect to $\phi$ gives
\begin{equation} \label{eq:field2}
\begin{aligned}
0 =& f'(X) \Big[ 2 \omega(\phi) \square \phi + \frac {d \omega(\phi)} {d \phi} \nabla_{\alpha} \phi \nabla^{\alpha} \phi \\
&+ R \frac {d F(\phi)} {d \phi} + \frac{ d V(\phi)} {d \phi} \Big] + 2 \omega(\phi) \nabla^{\alpha} \phi \nabla_{\alpha} f'(X),
\end{aligned}
\end{equation}
in {vacuum}. 

In general, several conditions are imposed on scalar-tensor dynamics to ensure a well-defined theory. {For example, demanding that the graviton carries a positive energy amounts to demonstrating that gravitational repulsion cannot occur. This can be achieved by linearizing the field equations about a perfect fluid background and expressing the resulting equations of motion in Poisson-like form \cite{fujii,damour92}. One may then read off the effective Newton `constant', which is required to be positive, from the coefficient of the mass density. This procedure however requires a non-vacuum perturbation analysis, and is challenging to carry out in general \cite{fel10}. Additionally, it is necessary for the astrophysical health of the theory that the kinetic energy of the scalar field be non-negative. This is guaranteed if the coefficient of $\square \phi$ within the field equation \eqref{eq:field2} is non-negative \cite{esp01}. Unfortunately, it is also non-trivial to check this condition in general because the Ricci scalar $R$ not only depends on $\square \phi$, as can be seen from equation \eqref{eq:field1}, but is also dynamical; Ricci modes can exist even in the $f(R)$ subcase (e.g. \cite{suv19}). A thorough check of these conditions involves detailed calculations on a case-by-case basis that will be conducted elsewhere. At least in the case of linear $f$ however, the aforementioned conditions reduce to the standard scalar-tensor ones $F(\phi) > 0$ and $2 F(\phi) \omega(\phi) + 3 \left[dF(\phi)/d\phi\right]^2 \geq 0$, respectively \cite{esp01}. These are automatically satisfied for the Brans-Dicke choices $F(\phi) = \phi$ and $\omega(\phi) \propto \phi^{-1}$ \cite{bd62}.}

For any $f$, however, an appealing feature of the theory described by \eqref{eq:genscalarten} {that can be proven without much difficulty is} that energy-momentum is conserved identically. Employing the Bianchi identities $\nabla_{\mu} \left( R_{\mu \nu} - \tfrac{1}{2} g_{\mu \nu} R \right) = 0$ and $\left( \square \nabla_{\nu} - \nabla_{\nu} \square \right) Z = R_{\mu \nu} \nabla^{\mu} Z$, the first of which is familiar from GR while the second is valid for any function $Z$ \cite{bianchi}, some extensive though not particularly difficult algebra shows that applying a contravariant divergence to the right-hand side of \eqref{eq:field1} produces a sequence of terms which vanish identically when equation \eqref{eq:field2} is used. As such, for the non-vacuum case where a stress-energy tensor $T_{\mu \nu}$ occupies the left-hand side of \eqref{eq:field1}, geometric identities give $\nabla^{\mu} T_{\mu \nu} = 0$ exactly, as in the pure $f(R)$ and scalar-tensor cases \cite{bianchi2}.

\section{Constructing a solution to the inverse problem}
In the case of pure $f(R)$ gravity, families of functions $f$ can be constructed such that any metric $g$ with constant scalar curvature, $R_{0}$, can be admitted as an exact solution. For example, if $f$ has a critical point at $R_{0}$ and also happens to vanish there (i.e., $f=0$ is a local extremum at the point $R_{0}$), the equations of motion are necessarily satisfied for any metric $g$ which has $R = R_{0}$. One such theory in this class is the Starobinsky-like quadratic theory with $f(R) = \left( R - R_{0} \right)^{2}$ \cite{quadratic}, for example. Therefore, in the case of constant-scalar-curvature (though not necessarily Einstein) spacetimes, certain $f(R)$ theories are already examples of solutions to the inverse problem\footnote{This implies that given \emph{any} metric $g$, the conformal metric $e^{2 \varphi} g$ for conformal factor $\varphi$ is a solution to \emph{some} $f(R)$ theory provided that the factor $\varphi$ is chosen such that the conformal scalar curvature is constant; mathematically, this requires the existence of a solution to the Yamabe problem \cite{yamabe1,yamabe2}. As such, practically any conceivable causal structure can arise in \emph{some} $f(R)$ theory \cite{suvm16}, because a metric conformally related to \emph{any} given metric can be admitted as an exact solution.}. While this is a somewhat trivial observation, it has, to our knowledge, not been expressly detailed elsewhere.

A similar but more extensive phenomenon to that described above exists in the generalized theories associated with the action \eqref{eq:genscalarten}. If the scalar field counterbalances the Ricci curvature in some precise way, the function $f$ within \eqref{eq:genscalarten} can be chosen to vanish at a realizable local extremum. As in the case of $f(R)$ gravity, this implies that, given any reasonable metric $g$, there exists a family of mixed scalar-$f(R)$ theories admitting that particular $g$ as an exact solution. 

To see this explicitly suppose that, for a given $g$ (reconstructed from astrophysical data, for instance), the scalar field $\phi$ solves {the \emph{kinematic constraint equation},}
\begin{equation} \label{eq:constraint}
X \equiv F(\phi) R + V(\phi) - \omega(\phi) \nabla_{\alpha} \phi \nabla^{\alpha} \phi = X_{0},
\end{equation}
{for some constant $X_{0}$.} If the function $f$ has a critical point at $X_{0}$ and also vanishes there, then the field equations are necessarily satisfied for this combination of $g$ and $\phi$, as each term within \eqref{eq:field1} and \eqref{eq:field2} can be seen to vanish. This means that, provided the scalar field can be chosen such that $\nabla_{\mu} X = 0$, there exists a function $f$ [e.g., $f(X) = (X-X_{0})^2$ for some $X_{0}$] for which some particular (though arbitrary) $g$ is an exact, vacuum solution to the theory governed by \eqref{eq:genscalarten}. In fact, there are infinitely many such functions. If we consider only those $f$ that are analytic, then the most general such $f$ can be represented as a power series, viz. $f(X) = \sum_{k \geq 2} a_{k} \left(X - X_{0} \right)^k$ for arbitrary coefficients $a_{k}$. Allowing for non-analytic $f$ further widens the class of suitable functions (see the example given in the next section).

In short, the main result of this Letter is that, for any given metric $g$, if
\begin{itemize}
\item[i)]{a scalar field $\phi$ can be chosen such that $X = X_{0}$ for some constant $X_{0}$, and}
\item[ii)]{the function $f$ satisfies $f(X_{0}) = f'(X_{0}) = 0$,}
\end{itemize}
then $g$ is a solution to the field equations \eqref{eq:field1} and \eqref{eq:field2} for the gravitational action \eqref{eq:genscalarten}. Note that these conditions are sufficient but not necessary; the Kerr metric, for example, can be an exact solution even when condition ii) is not satisfied \cite{psal08}.

It is important to note that we do not comment here on the physical viability or otherwise of such theories. Indeed, further analysis, beyond the scope of this Letter, is required to determine whether there exists members of the class constructed above that can accommodate existing (and upcoming) astrophysical experiments. For example, there may be no such $f$ which simultaneously satisfies the above and passes Solar System \cite{will18} and/or strong-field \cite{berti15} tests, even with screening mechanisms \cite{screen,brax13}. However, the exact conservation of energy-momentum hints that this may be possible.

\section{An example: parameterized black hole geometries}
Various techniques based on electromagnetic \cite{joh16,bam17} and gravitational-wave \cite{fran19,vol20} observations allow one to, with varying degrees of precision, identify the local spacetime geometry surrounding a monitored (usually compact) object. However, especially in the gravitational case, these tests inherently presuppose a particular set of field equations. Radiation of any sort saps energy from the system, and backreaction cannot be self-consistently accounted for without some overarching equations of motion. Backreaction effects are negligible in many cases of course, though those tests which involve oscillations or violent outbursts of compact objects may be sensitive to the particulars of the gravitational action \cite{bar08,suv19}. Metric reconstruction techniques, which use some parameterized scheme in lieu of an exact theory, are therefore limited in their validity to some degree \cite{kim20}. The recipe given in the previous section essentially allows one to build a theory around a given metric, which allows for a potential resolution to this problem. 

In this section, we show how one may tailor a particular theory of gravity to a given family of parameterized black holes, such as those considered in Refs. \cite{nkerr1,nkerr2,nkerr3,contfrac}. For demonstration purposes, suppose that astrophysical data implied that black holes were described by a simple generalization of the Kerr metric whose line element, in Boyer-Lindquist coordinates $(t,r,\theta,\varphi)$, reads
\begin{equation} \label{eq:defmet}
\begin{aligned}
\hspace{-0.4cm}ds^2 =& \frac {a^2 \sin^2\theta -\Delta} {\Sigma} dt^2 - \frac {2 a \sin^2\theta \left( a^2 + r^2 - \Delta\right)} {\Sigma} dt d \varphi \\
&+ \frac {\Sigma} {\Delta} dr^2 + \Sigma d\theta^2 + \frac { \left(a^2 + r^2 \right)^2 - a^2 \sin^2\theta \Delta} {\csc^2\theta \Sigma} d \varphi^2,
\end{aligned}
\end{equation}
where $\Delta = r^2 - 2 M r + a^2 + \epsilon M^3 /r$ and $\Sigma = r^2 + a^2 \cos^2\theta$. In expression \eqref{eq:defmet}, $M$ and $a$ denote the mass and spin of the black hole, respectively, while $\epsilon$ is a dimensionless `hair'. The metric \eqref{eq:defmet} admits an (outer) event horizon at the largest positive root of $\Delta = 0$, which occurs near the Kerr value for sufficiently small $\epsilon$, viz. $r \approx M + \sqrt{M^2 - a^2} + \mathcal{O}(\epsilon)$. 

The geometry described by \eqref{eq:defmet} represents a generalization of the Kerr spacetime with several desirable properties. Most notably, 1) the metric is asymptotically flat, 2) many post-Newtonian constraints \cite{will18} are automatically satisfied due to the absence of quadratic terms in the static limit $a =0$, and 3) the metric coefficients are algebraically simple, so that astrophysical tests involving electromagnetic data analysis are numerically easy to handle. The metric \eqref{eq:defmet} is a member of those considered in Ref. \cite{contfrac}, for instance.

{Consider, for example, the mixed scalar-$f(R)$ theory described by the action \eqref{eq:genscalarten} with $f(X) = X^{1+ \delta}$ for any $\delta > 0$. Such theories represent a generalization of the $f(R) = R^{1+\delta}$ theories studied by Clifton and Barrow \cite{clifbar}, which are known to admit non-Kerr solutions \cite{top2}. Note that this function $f$ is not analytic for $\delta \notin \mathbb{Z}$}, though $f=0$ is {still} a critical point at $X=0$ for any $\delta > 0$. The generalized theory \eqref{eq:genscalarten} therefore admits the metric \eqref{eq:defmet} as an exact solution, provided that the scalar field $\phi$ solves the {kinematic constraint equation \eqref{eq:constraint} with $X_{0} = 0$, i.e.,}
\begin{equation} \label{eq:constraint2}
0 = F(\phi) R + V(\phi) - \omega(\phi) \nabla_{\alpha} \phi \nabla^{\alpha} \phi.
\end{equation}
For the metric \eqref{eq:defmet} we find $R = - 2 M^3 \epsilon / \left(r^3 \Sigma \right)$, {which interestingly implies that we need only consider a radial scalar field when working with the Brans-Dicke choices $F(\phi) = \phi$, $V = 0$, and $\omega(\phi) = \phi^{-1}$ in the following sense. Taking a time- and azimuth-independent scalar field $\phi(r,\theta)$ so as to respect the Killing symmetries of the metric, we have that $\nabla_{\alpha} \phi \nabla^{\alpha} \phi = g^{rr} \left(\phi_{,r} \right)^2 + g^{\theta \theta} \left(\phi_{,\theta} \right)^2$. Since the contravariant components $g^{rr}$ and $g^{\theta \theta}$ are proportional to $\Sigma^{-1}$, the angular terms in \eqref{eq:constraint2} cancel out. It can be formally shown then that equation \eqref{eq:constraint2} implies $\phi_{,\theta} = 0$ when imposing periodic boundary conditions on $\theta$, i.e., demanding $\phi(r,\pi) = \phi(r,0)$ forces a radial scalar field. For a more general stationary metric or for a theory with a more involved scalar sector, however, the scalar field will necessarily depend on $\theta$.}

\begin{figure}[h]
\centerline{\includegraphics[width=0.5\textwidth]{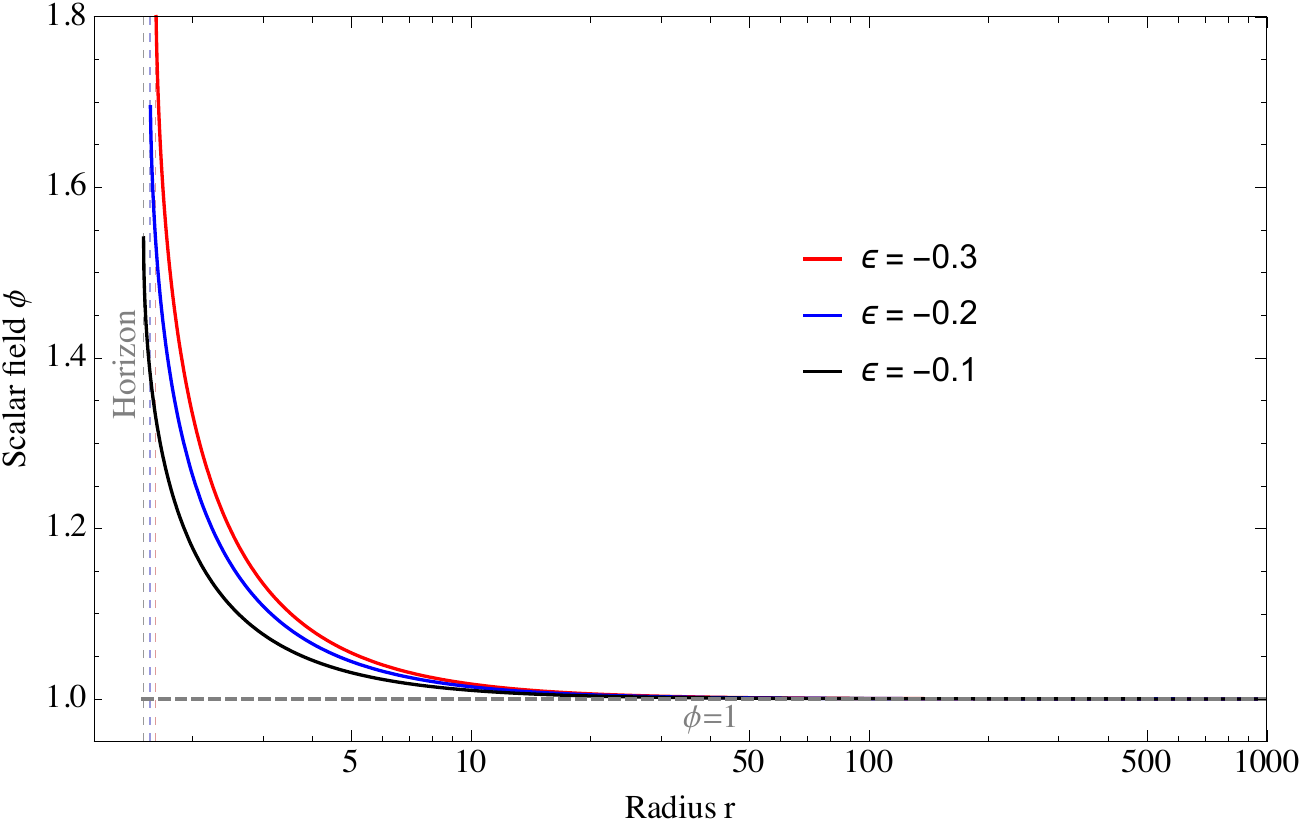}}
\caption{Radial scalar-field solutions to \eqref{eq:defcon} for $M=1$ and $a=0.9$, with $\epsilon = -0.1$ (black curve), $\epsilon = -0.2$ (blue curve), and $\epsilon = -0.3$ (red curve). The horizon in each case is shown by a vertical dashed line. \label{fig:fig1}}
\end{figure}

For simplicity, we therefore make the Brans-Dicke choices detailed above, though more complicated examples can be readily designed. In this case, equation \eqref{eq:constraint2} reduces to the simple form
\begin{equation} \label{eq:defcon}
0 = 2 M^3 \epsilon \phi(r)^2 + r^3 \Delta(r) \left[ \frac {d \phi(r)} {d r} \right]^2 .
\end{equation}
In general, there exists a well-behaved solution for $\phi$ {(i.e., one that is smooth, positive, and has non-negative kinetic energy; see Sec. 2)} to the constraint equation \eqref{eq:defcon} for a wide range of $\epsilon \leq 0$. Figure \ref{fig:fig1} shows numerical solutions to \eqref{eq:defcon} for $M=1$ and $a=0.9$ subject to the boundary condition $\underset{r \rightarrow \infty}{\lim} \phi(r) = 1$, which forces $\phi$ to asymptote towards the Newtonian {[i.e., $F(\phi) \rightarrow 1$]} value at large radii. Figure \ref{fig:fig1} illustrates that the scalar hair induced by the non-Kerr parameter $\epsilon$ is rather short-ranged, as $\phi \approx 1$ to within $1\%$ already for $r \gtrsim 30$ for all considered values of $\epsilon$. This particular field $\phi$, for which the metric \eqref{eq:defmet} is an exact solution to the theory \eqref{eq:genscalarten} for $f(X) = X^{1+ \delta}$, therefore appears to be well behaved and physically reasonable. For vanishing $\epsilon$ we find that $\phi$ is everywhere constant, as expected, since the Kerr metric is Ricci-flat and equation \eqref{eq:defcon} reduces to $d \phi / d r = 0$.

\section{Discussion}
In this Letter, a method is presented to build a covariant, Lagrangian theory of gravity around a pre-specified spacetime metric; in other words, a particular solution to the inverse problem in gravitation is found. Given some metric $g$, we show that a function $f$ and scalar field $\phi$ can often be found [so long as a solution to the kinematic constraint equation \eqref{eq:constraint} exists] such that $g$ is an exact solution to the mixed scalar-$f(R)$ theory governed by the action \eqref{eq:genscalarten}. For the particular case of $f(X) = X^{1+\delta}$ for $\delta > 0$, we found that a parametrically-deformed Kerr metric \eqref{eq:defmet} (cf. Ref. \cite{contfrac}) is an exact solution to the field equations \eqref{eq:field1} and \eqref{eq:field2}, provided that the scalar field satisfies the kinematic constraint \eqref{eq:defcon}. Solutions to equation \eqref{eq:defcon} are shown in Fig. \ref{fig:fig1} for a variety of black hole parameters. In all cases considered, the scalar field $\phi$ is short ranged, well behaved, and asymptotes to the Newtonian value $\phi_{\infty} = 1$, as expected of physical black hole geometries. Despite a number of attractive features, important questions remain about whether or not the theories considered herein are compatible with astrophysical constraints \cite{will18,berti15} {(see also Sec. 2)}. A thorough investigation will be conducted elsewhere.

One of the major benefits of the construction detailed herein is that gravitational perturbations of a given spacetime can be studied self-consistently. Given a solution to the equations of motion \eqref{eq:field1} and \eqref{eq:field2} [such as \eqref{eq:defmet}, for instance], a perturbation, encapsulated by the Eulerian scheme $g \rightarrow g + \delta g$ and $\phi \rightarrow \phi + \delta \phi$, can be introduced to deduce stability \cite{bhstab} and characterize any resulting gravitational radiation. {This may  lead to improvements in models of EMRIs \cite{nkerr1,emri2} or quasi-normal ringing \cite{vol20,kim20} that employ a parameterized non-Kerr object, since its response to astrophysical disturbances can be studied exactly in the theory described by \eqref{eq:genscalarten}.}

Some philosophical curiosities arise by noting that the approach presented here involves the construction of \emph{vacuum} solutions. Since the seed metric could arise as a matter-filled solution in GR (for example), this implies that vacuum gravitational fields in the theory governed by expression \eqref{eq:genscalarten} can imitate the gravitational fields of material bodies in a different theory. In this way, the gravitational field within and surrounding a star, for instance, could be mimicked by that of a vacuum object in the theory \eqref{eq:genscalarten}. This raises the interesting possibility of `gravitational doppelg{\"a}ngers'. Some examples of this phenomenon are already familiar from the literature; for instance, it is known that the electrovacuum Kerr-Newman metric arises as a pure vacuum solution in some modified theories of gravity \cite{top2,ransum}.

{Finally, it is worth noting that the general approach to the inverse problem considered here is not unique to scalar-tensor gravities. One could, for example, envision a mixed vector-tensor gravity, such that (functions of) gauge fields $A_{\mu}$ enter into the argument of $f$ instead of the scalar field $\phi$, where a similar constraint equation to that of \eqref{eq:constraint} may be imposed to provide a solution to the inverse problem. A larger solution space can be built in this way. Moreover, another avenue that would be worth considering involves finding a general solution for the inverse Kerr problem \cite{psal08}; one may attempt to map out the space of theories that abide by the classical no-hair theorems using such an approach (see also Ref. \cite{suv20}).
}

\section*{Acknowledgements}
Thanks are due to Sebastian V{\"o}lkel and Sourabh Nampalliwar for comments made on an earlier version of this manuscript, {and to the anonymous referee for providing helpful feedback which significantly improved the quality of this work.} 

\section*{Declarations} 
This work was supported financially by the Alexander von Humboldt Foundation. There are no conflicts of interest to report. Data will be made available upon request.

\end{document}